\documentclass[amsmath,amssymb,floatfix,pra,epsf,twocolumn,superscriptaddress]{revtex4-1}

\usepackage{amsmath,amssymb,bm,graphicx,url,epsfig}
\usepackage[ansinew]{inputenc}
\usepackage{bm}
\usepackage{color}

\newcommand{\be}{\begin{equation}}
\newcommand{\ee}{\end{equation}}
\newcommand{\bes}{\begin{equation}\begin{split}}
\newcommand{\ees}{\end{split}\end{equation}}
\newcommand{\bea}{\begin{eqnarray}}
\newcommand{\eea}{\end{eqnarray}}

\def\beq{\begin{equation}}
\def\eeq{\end{equation}}
\def\bea{\begin{eqnarray}}
\def\eea{\end{eqnarray}}
\begin{document}

\title{Comment on ``Kinetic theory for a mobile impurity \\in a degenerate Tonks-Girardeau gas''}


\author{Michael Schecter}

\affiliation{School of Physics and Astronomy, University of Minnesota, Minneapolis, Minnesota 55455, USA}

\author{Dimitri M. Gangardt}

\affiliation{School of Physics and Astronomy, University of Birmingham, B15 2TT, United Kingdom}

\author{Alex Kamenev}

\affiliation{School of Physics and Astronomy, University of Minnesota, Minneapolis, Minnesota 55455, USA}

\affiliation{William I. Fine Theoretical Physics Institute, University of Minnesota, Minneapolis, Minnesota 55455, USA}

%

\maketitle

In a recent paper  \cite{Lancaster}  Gamayun, Lychkovskiy, and Cheianov studied the dynamics of a mobile impurity weakly coupled to a one-dimensional Tonks-Girardeau gas of strongly interacting bosons.  Employing the Boltzmann equation approach,  they arrived at the following conclusions: (i) a light impurity, being accelerated by a constant force, $F$, does {\em not} exhibit Bloch oscillations, which were  predicted and studied in Refs.~\cite{Gang_Kam,Schec1}; (ii) a heavy impurity does undergo Bloch oscillations, accompanied by a drift with the velocity $v_D \propto \sqrt{F}$.  

In this comment we argue that the result (i) is  an artifact of the classical Boltzmann approximation. The latter misses the formation of the (quasi) {\em bound-state} between the impurity and a hole. Its dispersion relation $E_b(P,\rho)$ is a {\em smooth} periodic function of momentum $P$ with the period $2k_F =2\pi\hbar \rho$, where $\rho$ is a density of the host gas.  Being accelerated by a small force, such a bound-state  exhibits Bloch oscillations superimposed with the drift velocity $v_D=\mu F$.  The mobility $\mu$ may be expressed {\em exactly} \cite{Schec1}  in terms of $E_b(P,\rho)$.  
Result (ii), while not valid at exponentially small forces, indeed reflects an interesting intermediate-force behavior. 

The origin of Bloch oscillations is most transparent for a weakly interacting Bose gas, described by the Gross-Pitaevskii (GP) equation. Its solution reveals that a repulsive impurity binds to a dark soliton -- a region of depleted host gas. The resulting composite object  (the ``depleton'') has a periodic dispersion curve and thus exhibits Bloch oscillations, if a sufficiently small    
force is applied to the impurity. In a strongly interacting Bose gas the GP approach is not applicable, but the bound-state formation still takes place. To illustrate this phenomenon, one may represent the Tonks-Girardeau gas of $N$ hard-core bosons  by  free fermions created by $c^\dagger_p$, weakly coupled to a quantum impurity with the coordinate $x_i$ through the density-density interaction $(\hbar=1)$:  
\begin{equation}
                                                                       \label{eq:H}
\hat H=-\frac{1}{2m_i}\,\frac{\partial^2}{\partial x_i^2} +\sum_p \frac{p^2}{2m_h} c^\dagger_p c_p + 
\gamma\frac{\rho^2}{m_h N} \sum_{p,q} c^\dagger_pc_{p+q} e^{iqx_i}\,
\end{equation}
where $m_h$ is the mass of host particles, $m_i$ is the impurity mass  and $0<\gamma \ll 1$ is a dimensionless coupling constant.

\begin{figure}

\includegraphics[width=.98\columnwidth]{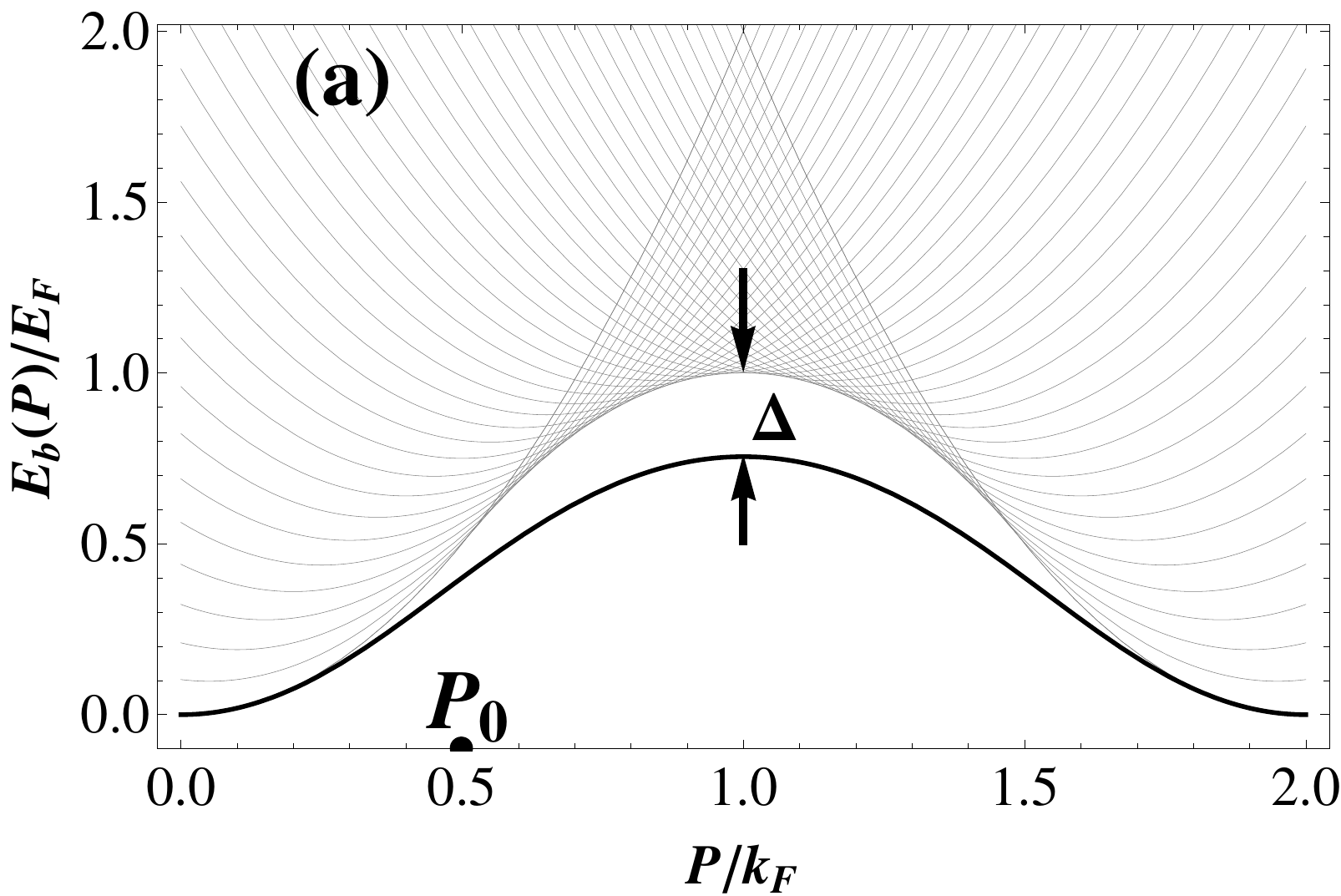}
\includegraphics[width=.98\columnwidth]{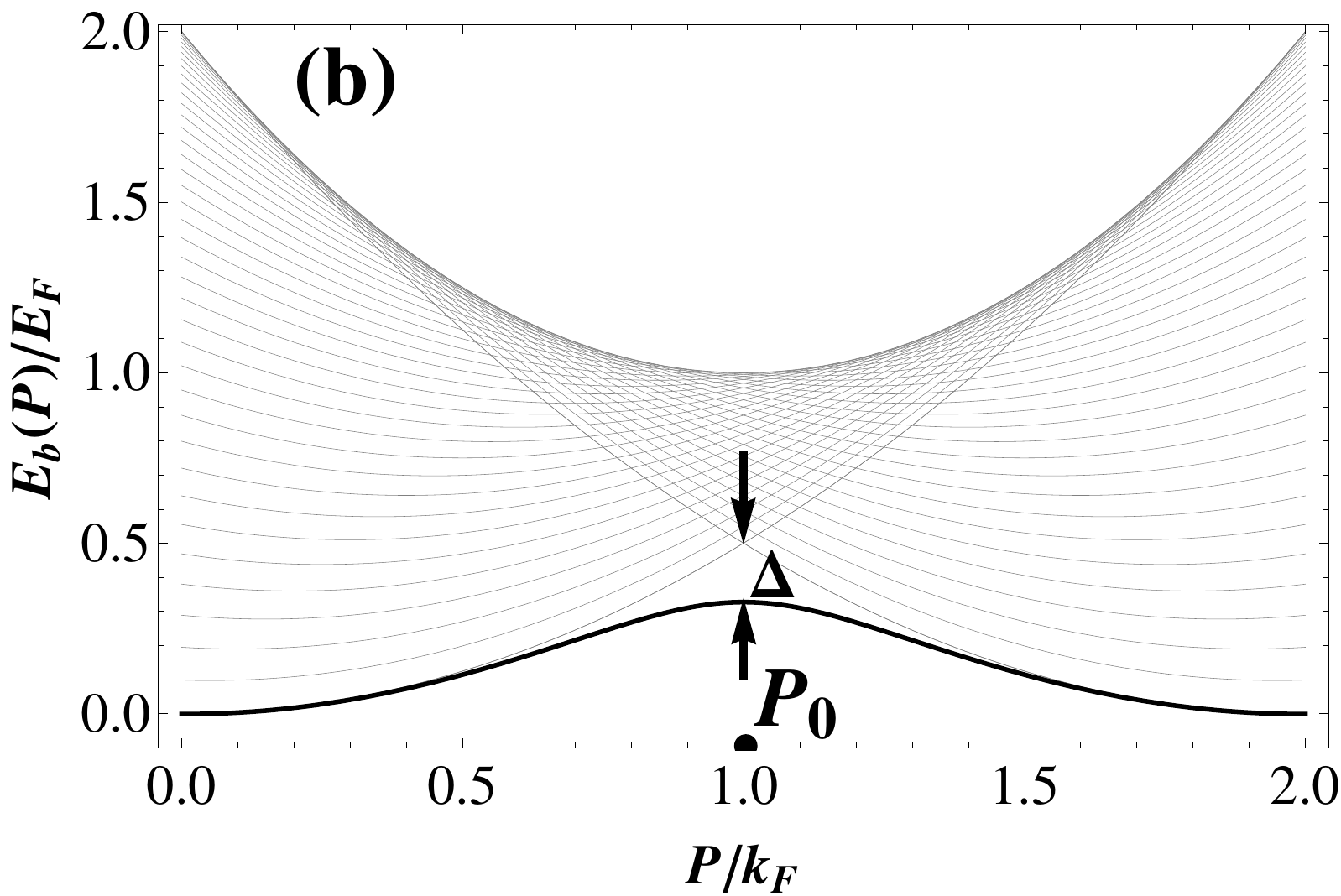}
\caption{(Color online) The bound-state $E_b(P)$, Eq.~(\ref{eq:integral}), (thick black line) and scattering continuum $\frac{(P- k)^2}{2m_i}+E_h(k)$ for a set of $k$ (thin gray lines) for the light impurity $\eta =1/2$ (a) and heavy impurity $\eta=2$ (b). In both cases $\gamma=0.7$.   }
\label{fig:fig1}
\end{figure}

Consider a state of the system with total momentum $P>0$. If $P<P_0\equiv\mathrm{min}\{m_iv_F,k_F\}$, the low energy states are those where most of the momentum is carried by the impurity. Indeed, the impurity kinetic energy $P^2/(2m_i)$ is less than that of soft particle-hole excitations above the Fermi sea $\sim v_F P$. In the opposite limit $P>P_0$ the low energy states are those where {\em hole excitations} carry a significant fraction of the entire momentum $P$. The many-body ground state  adiabatically connects between these two limits, signaling  strong impurity-hole hybridization at $P>P_0$. Indeed, 
consider a subspace of the full many-body space, which contains a single hole excitation with momentum $0<k<2k_F$ in addition to the impurity with momentum $P-k$ (this restriction is justified in the limit $\gamma\ll 1$).  The basis vectors of this 
subspace are  
\begin{equation}
                                                                     \label{eq:psi}
|k;P\rangle=  e^{i(P-k)x_i} c^\dagger_{k_F}c_{k_F- k}|\mbox{Fermi Sea}\rangle\,. 
\end{equation}
The corresponding Schr\"odinger equation $\sum_{k'} \langle k;P| \hat H |k';P\rangle \psi_P(k') =E\psi_P(k)$ takes the form 
of the two-particle problem  with the {\em attractive} delta-interaction (formally the attraction arises from anti-commuting the fermionic operators in the last term in Eq.~(\ref{eq:H})),
 \begin{equation}
                                                                      \label{eq:Schrodinger}
\left[\frac{(P- k)^2}{2m_i}+\!E_h(k)\right] \psi_P(k) - \frac{\gamma\rho}{m_h}\!\! \int\limits_{0}^{2k_F}\! \frac{dk'}{2\pi}\,\psi_P(k')\!=\!E\psi_P(k),  
\end{equation} 
where $E_h(k)=v_Fk-k^2/(2m_h)$ is the hole kinetic energy (we measure $E$ relative to $NE_F/3+\gamma\rho^2/m_h$). This problem admits a unique bound-state solution, whose energy $E=E_b(P)$ is found from the integral equation  
 \begin{equation}
                                                                      \label{eq:integral}
\int\limits_{0}^{2k_F} \frac{dk'}{\frac{(P- k')^2}{2m_i}+E_h(k') -E_b(P)} = \frac{2\pi m_h}{\gamma\rho}\,. 
\end{equation} 
Its solution represents a non-perturbative correction to the bare impurity dispersion and is completely missed in the Boltzmann equation treatment. We plot $E_b(P)$, along with the continuum of the scattering states, in Fig.~\ref{fig:fig1} for the case of light, $\eta=m_i/m_h<1$, and heavy, $ \eta > 1$ impurity. The gap $\Delta$ between the bound-state and the continuum is found to be $\Delta/E_F\sim \gamma^2 \eta/(1-\eta)$ for $\eta < 1-\gamma/\pi^2$ and $P_0\leq P$, while 
$\Delta/E_F\sim \exp\{-\pi^2(\eta-1)/\eta\gamma\}$ for $\eta >1+\gamma/\pi^2$. For an almost equal mass case $|1-\eta|<\gamma/\pi^2$, one finds  $\Delta/E_F\sim \gamma$. We also note that for $\eta=1$, integrability of Eq.~(\ref{eq:H}) allows access to the exact many-body ground state energy \cite{Lamacraft} $E_0(P\sim k_F)=E_F-\frac{2\pi^2}{3\gamma}\frac{(P-k_F)^2}{2m_h}$. Remarkably, as one may verify from Eq.~(\ref{eq:integral}), $E_b(P\sim k_F)+\gamma\rho^2/m_h=E_0(P\sim k_F)$ for $\gamma\ll1$, justifying our Hilbert space truncation.

The hard gap between the bound-state and the continuum is an artifact of  restricting  the particle in Eq.~(\ref{eq:psi}) to be created right at the Fermi momentum $k_F$. Allowing for slight deviation $c^\dagger_{k_F} \to 
c^\dagger_{k_F+p}$, introduces interaction of the bound-state with low energy, $\sim v_Fp$, excitations.  It is known \cite{Kam_Glazman,Glazman,Lamacraft} that such interaction transforms the bound-state into the {\em quasi} bound-state with the power-law (instead of the pole) correlation  function. These low energy excitations are responsible for radiation losses and thus for linear mobility $\mu$. They do not, however, destroy the quasi bound-state and associated Bloch oscillations at small applied force.

The Bloch oscillations are destroyed if a large enough force $F>F_{\mathrm{max}}$ is applied to the impurity. The physics of this 
process is that of the Landau-Zener transition between the bound-state and the continuum at $P\approx P_0=k_F\,\mathrm{min}\{\eta,1\}$. One may thus estimate the crossover force as $F_{\mathrm{max}}\sim \Delta^2/v$, where 
$v=v_F\,\mathrm{min}\{1,1/\eta\}$. This leads to the following estimate for the maximal force, preserving (nearly) adiabatic bound-state dynamics

\begin{equation}
						\label{eq:F_max}
F_{\mathrm{max}}\propto \frac{ k_{F}^{3}}{m_h}\begin{cases}
\left(\frac{\gamma^2\eta}{1-\eta}\right)^2, & \eta<1-{\gamma\over\pi^2};\\
\eta\left(\frac{\eta-1}{\eta}\right)^2\mathrm{e}^{-\frac{2\pi^2(\eta-1)}{\gamma\eta}}, & \eta>1+{\gamma\over\pi^2},\end{cases}
\end{equation}
while for $|1-\eta|<\gamma/\pi^2$, one finds $F_{\mathrm{max}}\propto k_F^3\gamma^2/m_h$. 
%
For $F<F_{\mathrm{max}}$ both light and heavy impurities exhibit Bloch oscillations along with the drift \cite{Schec1},  whose velocity scales linearly with the force $v_D=\mu F$.  

  In Refs.~\cite{Lamacraft,Schec2} it was  shown that for a heavy impurity {\em away} from the Tonks-Girardeau limit, there exists a phase transition at a critical value of the impurity mass: for $m_i<M_c$ the ground-state is a smooth function of momentum, while for $m_i>M_c$ the ground-state exhibits a cusp singularity at momenta $P=(1+2n)k_F$ for integer $n$ (in the Tonks-Girardeau limit $M_c\to \infty$).  In the latter case the impurity
``overshoots'' the intersection points at $P=(1+2n)k_F$ and has to emit phonons to reach the ground state.  This leads to 
an enhanced dissipation \cite{Schec2} and thus to super-linear drift velocity  
\begin{equation}
                                  \label{eq:power}
v_D\propto F^{1/\left(1+\alpha\right)},
\end{equation}
where $\alpha\approx 2K-1$ for $\gamma\ll 1$ and $K$ is the Luttinger parameter of the host.

Notice that in the Tonks-Girardeau limit  the validity of  the $v_{D}=\mu F$ response for $\eta>1$ is limited to an exponentially small force (\ref{eq:F_max}). This scale originates from the exponentially narrow region of momenta,  where the bound-state exhibits the avoided crossing behavior, Fig.~\ref{fig:fig1}(b). For $F>F_{\mathrm{max}}$ the impurity 
overshoots the avoided crossing and follows the ``wrong'' parabola before emitting phonons and returning to the ground state. Thus, for $F>F_{\mathrm{max}}$ one may apply Eq.~(\ref{eq:power}) with $K=1$ -- appropriate for the Tonks gas. This leads to $v_D\propto \sqrt{F}$, in full agreement with  Ref.~\cite{Lancaster}. An important extension  of  Ref.~\cite{Lancaster} is that the super-linear drift (\ref{eq:power}) is to be expected for moderately heavy impurities
$m_h<m_i<M_c$ in an intermediate range of forces where the linear mobility  $v_{D}=\mu F$ is inapplicable.  

M. S. and A. K. were supported by DOE Contract No. DE-FG02-08ER46482. D. M. G. acknowledges support by the EPSRC.

\end{document}